\documentclass[11pt]{article}
\usepackage{amssymb}
\usepackage{amsmath}
\usepackage{graphicx}
\usepackage{color}

\def\beq{\begin{eqnarray}}    
\def\eeq{\end{eqnarray}}      

\def\beq{\begin{eqnarray}}    
\def\eeq{\end{eqnarray}}      

\newcommand{\OMo}{\Omega_{m}^0}

\newcommand{\OLo}{\Omega_{\Lambda}^0}

\newcommand{\rmo}{\rho_{m}^0}

\newcommand{\rM}{\rho_m}
\newcommand{\rmr}{\rho_m}
\newcommand{\pmr}{p_m}

\newcommand{\rL}{\rho_{\CC}}

\newcommand{\rLo}{\rho_{\CC}^0}

\newcommand{\CC}{\Lambda}

\newcommand{\num}{\nu_m}
\newcommand{\nuL}{\nu_\CC}
\newcommand{\nuG}{\nu_{\rm G}}

\newcommand{\nueff}{\nu_{\rm eff}}

\newcommand{\LQCD}{\Lambda_{\rm QCD}}
\newcommand{\OMB}{\Omega_B}
\newcommand{\OMBo}{\Omega^0_B}
\newcommand{\ODM}{\Omega_{\rm DM}}
\newcommand{\ODMo}{\Omega^0_{\rm DM}}
\newcommand{\mupe}{\mu_{\rm pe}}

\newcommand{\nuX}{\nu_{X}}
\newcommand{\nuB}{\nu_{B}}

\newcommand{\be}{\begin{equation}}
\newcommand{\ee}{\end{equation}}

\begin{document}



 \hyphenation{cos-mo-lo-gi-cal
sig-ni-fi-cant par-ti-cu-lar}




\begin{center}
{\LARGE \textbf{Fundamental constants and cosmic vacuum: \\ the micro and macro connection}} \vskip 2mm

\vspace{0.8cm}

\textbf{\large Harald Fritzsch\,$^{a,c}$, and  Joan Sol\`{a}\,$^{b,c}$}

\vspace{0.5cm}

$^{a}$ Physik-Department, Universit\"at M\"unchen, D-80333 Munich,
Germany

\vspace{0.3cm}

$^{b}$ High Energy Physics Group, Dept. ECM and Institut de Ci{\`e}ncies del Cosmos\\
Univ. de Barcelona, Av. Diagonal 647, E-08028 Barcelona, Catalonia, Spain

\vspace{0.5cm}

$^{c}$ Institute for Advanced Study,
Nanyang Technological University, Singapore

\vspace{0.25cm}

E-mails: fritzsch@mppmu.mpg.de, sola@ecm.ub.edu \vskip2mm

\end{center}
\vskip 15mm

\begin{quotation}
\noindent {\large\it \underline{Abstract}}.\ \
The idea that the vacuum energy density $\rL$ could be time dependent is
a most reasonable one in the expanding Universe; in fact, much more
reasonable than just a rigid cosmological constant for the entire
cosmic history. Being $\rL=\rL(t)$ dynamical, it offers a
possibility to tackle the cosmological constant problem in its
various facets. Furthermore, for a long time (most prominently since
Dirac's first proposal on a time variable gravitational coupling)
the possibility that the fundamental ``constants'' of Nature are
slowly drifting with the cosmic expansion has been
continuously investigated. In the last two decades, and specially in recent
times, mounting experimental evidence attests that this
could be the case. In this paper, we consider the possibility that
these two groups of facts might be intimately connected, namely that
the observed acceleration of the Universe and the possible time
variation of the fundamental constants are two manifestations of the
same underlying dynamics. We call it: the ``micro and macro connection'', and on its basis we expect that the
cosmological term in Einstein's equations, Newton's coupling and the
masses of all the particles in the Universe, both the dark matter
particles and the ordinary baryons and leptons, should all drift
with the cosmic expansion. Here we discuss specific cosmological
models realizing such possibility
in a way that preserves the principle of covariance of General Relativity.

\end{quotation}
\vskip 8mm


\newpage

\vskip 6mm

 \noindent \section{Introduction}
 \label{Introduction}

The possibility that the constancy of the fundamental ``constants of
Nature'' could just be a particular appreciation of human beings,
who tend to compare the rhythm of natural phenomena with their
average lifetime, has been in the literature since long time ago.
The history traces back mainly to Dirac's large number hypothesis in
the thirties\,\cite{Dirac3738}, from which a time evolution of the
gravitational constant $G$ was suggested. The idea was disputed by
E. Teller\,\cite{Teller1948} on a geophysical basis and further
qualified by R.H. Dicke\cite{Dicke1957}. It also  triggered
subsequent speculations by G. Gamow\,\cite{Gamow1967} on the
possible variation of the fine structure constant. Despite the
initial difficulties, these seminal thoughts were a real spur to start
changing our minds on the supposedly imperturbable and rigid status
of the ``sacrosanct'' constants of Nature.

Since then the subject has been in continuous evolution and new
hints of experimental evidence have been piling up. For example, in
the last decade and a half or so \cite{Webb1999}
different sources of experimental information
coming from astrophysical observations using
absorption systems in the spectra of distant quasars have
monitored the possible time change of the fine structure constant
and started to provide  a body of detailed astronomical
evidence, or at least of suggestive possibilities. There emerged, for
instance, groups claiming to have discovered a positive effect at
the level of 4-5$\sigma$\,\cite{Webb2001Murphy2001,MurphyWebbFlambaum2003}, while other groups
cannot confirm this result\,\cite{Chand:2004ct,Srianand2007}. Apart from rolling in time, the couplings could also be rambling in space. In fact,
intriguing indications from quasar absorption lines suggesting a
space variation of the fine structure constant at the $\sim 4\sigma$
level have been put forward quite recently\,\cite{Flambaum2012}. Theoretical ideas have already been developed to try to explain this possibility, see e.g.\cite{BarrowMagueijo2014} (and references therein).
Similarly, experimental observations (once more using distant
quasars)  claim to have detected a significant time evolution of the
proton-electron mass ratio\,\cite{Reinhold06}.

Clearly, the time and space variation of the fundamental constants
is a very active field of theoretical and experimental research that
could eventually provide interesting surprises in the near future --
see e.g. \,\cite{FundamentalConstants,Barrow2009,Calmet2014,GarciaBerro2007} for reviews. Such variation of the constants suggest that basic quantities of the
standard model, such as the QCD scale parameter $\LQCD$, might not
be conserved in the course of the cosmological evolution\,\cite{CalmetFritzsch}. If so the
masses of the nucleons and of the atomic nuclei would be
time-evolving. Remarkably, this can be consistent with General
Relativity (GR) provided the vacuum energy density, $\rL=\CC/8\pi G$, is a dynamical quantity: $\rL=\rL(t)$ . This fact speaks up of the
possibility of the deep connection between the laws of the micro and
macro cosmos\,\cite{FritzschSola2012,FritzschSola2014,Sola2014}.

In an expanding Universe the idea that the cosmological term, $\CC$,
and Newton's gravitational coupling, $G$, could be variable with
time  can be viewed as a reasonable assumption. Historically, we have the well-known attempts by Jordan, Fierz and Brans-Dicke\,\cite{JFBD}, who first extended GR to accommodate variations in $G$.  For recent studies on testing $G$-variable theories through gamma ray bursts, see e.g. \cite{Capozziello}.  As for the $\CC$-term, it was
introduced by Einstein 98 years ago and was assumed to be
positive\,\cite{Einstein1917}. The reasons why Einstein introduced
that term are well-known\,\cite{CCP}\,\footnote{See e.g.
Ref.\,\cite{JSP-CCReview2013} for a recent review focusing on these
points.}, and at present are of mere historical interest. The
assumption of constancy is probably an approximation, certainly a good
one for a fraction of a Hubble time, but it is most likely a
temporary description of a true dynamical vacuum energy variable
that is evolving from the inflationary epoch to the present day. A
fundamental explanation of the so-called ``cosmological constant
problem''\,\cite{CCP,JSP-CCReview2013,Steven2015} demands to be much more open minded
as to the nature of the cosmological term. The final solution may well be much more flexible and ultimately point
once more towards its likely dynamical
character\,\cite{JSP-CCReview2013,SolGoReview2015}.

In this paper we will further dig out into the possible link between
the micro and macro cosmos, namely between the profound
interconnection that could exist between the possible time evolution
of the particle physics fundamental constants
and the associated time evolution of the gravitational parameters,
particularly Newton's gravitational coupling $G$ and the
cosmological term $\CC$.

\section{Time-evolving couplings and masses}\label{sect:couplings and masses}

When we consider that a fundamental ``constant'' of Nature could be slowly time-evolving, $f=f(t)$, it is natural to assume that the time scale of such evolution should be proportional to the rate of change of the scale factor of the cosmic expansion, i.e.  $\dot{f}/f\propto\dot{a}/a\equiv H$ (the Hubble rate). Thus, at present we expect that the cosmic drift rate of $f(t)$ is proportional to $H_0=1.0227\,h\times 10^{-10}\,{\rm
yr}^{-1}\,$, with $h\simeq 0.67$. Needless to say, this is only  a linear approximation to the dynamics of such time variation, which in practice could be more complicated. But as a first estimate for a quantity that is assumed to be essentially constant over a cosmic span of time should be reasonable.
On the other hand, the cosmic time
variation can be related with the corresponding redshift variation, as follows:
\begin{equation}\label{eq:dotffp}
\frac{\dot{f}}{f}\equiv
\frac{1}{f(t)}\frac{df(t)}{dt}=\frac{1}{f}\frac{df}{dz}\frac{dz}{da}\,\dot{a}=-(1+z)\,H(z)\,\frac{f'(z)}{f(z)}\,,
\end{equation}
where $z=(1-a)/a$ is the redshift, $a(t)$ is the scale factor
(normalized to $a(t_0)=1$ at present) and $H=\dot{a}/a$ is the
Hubble rate.  A cosmic time evolution of any parameter ${\cal
P}={\cal P}(t)$ (whether couplings, masses etc) is thus equivalently
described as a redshift dependence of that parameter, ${\cal
P}={\cal P}(z)$. We will use Eq.\,(\ref{eq:dotffp}) frequently in this work.

The variation of masses is, of course, not entirely independent of
the variation of couplings. This is clear if one thinks on radiative
corrections. If the fine structure constant $\alpha_{\rm
em}=\alpha_{\rm em}(t)$ can be (slowly) varying with cosmic time or
redshift, the masses of all nucleons can vary as well since the
interaction responsible for the variation of $\alpha_{\rm em}$
should couple radiatively to nucleons. As a result one expects the
proton and neutron masses to be time/redshift
dependent\,\cite{OlivePospelov2002}. In such context, the link
between the two time variations would still reside in the particle
physics world. However, the source of time/redshift evolution of
masses and couplings that we will address here is of different
nature. It has been proposed in \cite{FritzschSola2012}. It stems
from the cosmic evolution of the gravitational parameters $\CC$ and
$G$. In an expanding Universe there is no natural reason to expect
that the vacuum energy density remains rigid; and once this is
admitted the general covariance of the theory leads naturally
to a variable gravitational coupling.  It is a challenging
possibility that we will discuss here. But let us first briefly
recall a few milestones on the possible cosmic evolution of the
particle physics parameters.

\subsection{The Oklo phenomenon and other hints of $\alpha_{\rm em}$ time-variation}

Constraints on $\dot{\alpha}_{\rm em}/\alpha_{\rm
em}\equiv(1/\alpha_{\rm em})d\alpha_{\rm em}/dt$ can be derived from
limits on the position of nuclear resonances in natural fission
reactors that have been working for the last few billions years --
e.g. the so-called ``Oklo phenomenon''. It is related to the natural
fission reactor (the Oklo uranium mine in
Gabon)\,\cite{Shlyakhter1976,DamourDyson96,Davis2014}. It
operated nearly 2 billion years ago ($z\simeq 0.16$) for a period of some two hundred
thousand years at a power of $\sim 100$ Kw.  The fraction of
$^{235}U$ in the Oklo site has decreased since then with respect to
the current standard value, and this is interpreted as a proof of
the past existence of a spontaneous chain reaction.  The cross
section of the neutron capture depends on the energy of a resonance
at $E_r = 97.3$ meV. The uncertainty in the resonance energy,
$\delta E_r$, is set equal to $E^{\rm Oklo}-E_r^0$, where $E^{\rm
Oklo}$ is the value of the resonance during the Oklo phenomenon and
$E_r^0$ is the possibly different value taken today. From the mass
formula of heavy nuclei the change in resonance energy is related to
a possible change in $\alpha_{\rm em}$ through the Coulomb energy
contribution:
\begin{equation}\label{eq:Coulomb}
\delta E_r=-1.1\,\frac{\delta\alpha_{\rm em}}{\alpha_{\rm em}}\,{\rm
MeV}\,.
\end{equation}
From the estimates on $\delta E_r$ (ranging from a dozen meV to a
hundred MeV\,\cite{DamourDyson96,Davis2014}) one infers
from (\ref{eq:Coulomb}) a tight  bound on the fractional variation of the fine
structure constant\,\footnote{Uncorrelated with potential changes in the light quark masses\,\cite{Davis2014}.}, ranging from  $\left|\Delta{\alpha}_{\rm
em}/\alpha_{\rm em}\right|=\left(-0.9\rightarrow +1.1\right)\times 10^{-7}$ to the tightest one $\left|\Delta{\alpha}_{\rm
em}/\alpha_{\rm em}\right|=\left(-0.1\rightarrow+0.07\right)\times10^{-7}$ (cf. \cite{Davis2014}), corresponding to a an approximate time variation of order
$\left|\dot{\alpha}_{\rm em}/\alpha_{\rm em}\right|\lesssim
10^{-17}\,{\rm yr}^{-1}\,$ (assuming a
linear change in time) . This is competitive with the best bounds
from atomic clocks, see e.g. \,\cite{FundamentalConstants,Barrow2009,Calmet2014}.
Earlier experiments, e.g. those based on measuring the decay of
radio-isotopes in meteorites, furnished significant bounds of order
$\left|\Delta{\alpha}_{\rm em}/\alpha_{\rm em}\right|< 10^{-4}$, or
$\left|\dot{\alpha}_{\rm em}/\alpha_{\rm em}\right|<
10^{-13}$yr$^{-1}$, see\,\cite{Dyson1972}. Most data sets provide
some sensitivity in the more modest range $\left|\Delta{\alpha}_{\rm
em}/\alpha_{\rm em}\right|\sim 10^{-5}$.

Nowadays direct astrophysical observations are very helpful in this
respect. Observing the light emission from quasars and paying
attention to the absorption spectrum in interstellar clouds,  one
can monitor the possible redshift variation of $\alpha_{\rm em}$.
This has been done e.g. by the Keck  telescope, finding a $\sim 5\sigma$ effect
$\left|\Delta{\alpha}_{\rm em}/\alpha_{\rm em}\right|=\left(-0.543\pm 0.116\right)10^{-5}$\,\cite{MurphyWebbFlambaum2003,Murphy2004}. These positive results
are however disputed by VLT
observations\,\cite{Chand:2004ct,Srianand2007,Webb2011}. The tension between the
two might, however, be resolved by the aforementioned possible
spatial variation of $\alpha_{\rm em}$, as it can introduce
directional dependent changes in the look-back
time\,\cite{Webb2011,Flambaum2012}. Let us finally mention the recent claims on the tightest bounds at the level of $0.1$ ppm (parts-per-million), $\left|\Delta{\alpha}_{\rm em}/\alpha_{\rm em}\right|\sim 10^{-7}$\,\cite{Truppe2013},  extracted from accurate measurements of microwave frequency transitions of the CH molecule in the lab as compared to those measured from interstellar sources of CH in the Milky Way.  If confirmed, such bounds would be as tight as the original ones from the Oklo phenomenon.

\subsection{The cosmic time evolution of $\LQCD$ and $m_p$}

Strong interactions could also be involved in the cosmic time
evolution of the fundamental constants of particle physics. The most
relevant parameter in this case is the QCD scale parameter. It is
related to the strong coupling constant $\alpha_s=g_s^2/(4\pi)$ as
follows (at 1-loop order):
\begin{equation}\label{alphasLQCD}
\alpha_s(\mu_R)=\frac{2\pi}{{b}\,\ln{\left(\mu_R/\LQCD\right)}}\,.
\end{equation}
Here $b=11-2n_f/3$ is the 1-loop $\beta$-function coefficient, with
$n_f$ the number of quark flavors, and $\mu_R$ is the renormalization
point. The strong coupling ``runs'' with $\mu_R$, this is the
conventional running of the strong gauge coupling, but as we shall see other running scales are possible. The value of the
QCD scale parameter is of order $\LQCD={\cal O}(200)$ MeV. While
$\LQCD$ is fixed in the standard model, it  could change with the
cosmic expansion and hence with the cosmic time/redshift. In this
case $\alpha_s(\mu_R; t)$ would run both with the renormalization
scale $\mu_R$ and the cosmic time $t$. One can easily show from
(\ref{alphasLQCD}) that the relative cosmic variations of the two
QCD quantities are related (at one-loop) by:
\begin{equation}\label{eq:timealphaLQC}
\frac{1}{\alpha_s}\frac{d\alpha_s(\mu_R;\xi)}{d\xi}=\frac{1}{\ln{\left(\mu_R/\LQCD(\xi)\right)}}\,\left[\frac{1}{\LQCD}\,\frac{d{\Lambda}_{\rm
QCD}(\xi)}{d\xi}\right]\,,
\end{equation}
where $\xi$ is any dynamical cosmic variable, whether the cosmic time $t$ or any quantity depending on it, e.g. the Hubble rate $H=H(t)$. In such dynamical framework the nucleon masses and the masses of the
atomic nuclei would change with time accordingly. Indeed, take the
current value of the proton mass, which is known with high precision:
$m_p^0=938.272046(21)$ MeV\,\cite{Mohr2012}. It can be computed in
QCD using  $\LQCD$, the quarks masses and the (small)
electromagnetic contribution:
\begin{eqnarray}\label{eq:ProtonMass}
m_p &=&c_{\rm QCD}\LQCD+c_u\,m_u+c_d\,m_d+c_s\,m_s+c_{\rm
em}\LQCD\,,
\end{eqnarray}
where the bulk effect ($860$ MeV) comes from the first $\LQCD$ term
on its \textit{r.h.s.}. Therefore, if $\LQCD=\LQCD(t)$ we must
necessarily have $m_p=m_p(t)$ as well, and since the
$\LQCD$-component dominates we find that the respective time
variations satisfy
\begin{eqnarray}\label{eq:VariationProtonMass}
\frac{\Delta m_p}{m_p}\simeq \frac{\Delta\LQCD}{\LQCD}\,.
\end{eqnarray}

In practice one considers the mass ratio involving the electron's mass:
\begin{equation}\label{eq:mupe}
\mupe\equiv\frac{m_p}{m_e}\,.
\end{equation}
This ratio is also known with high accuracy:
$\mupe=1836.15267245(75)$\,\cite{Mohr2012}. Since a change of
$\LQCD$ would not affect the electron mass, the mass ratio
(\ref{eq:mupe}) would change during the cosmological evolution in a
similar way as in (\ref{eq:VariationProtonMass}).

The spectrum of the hydrogen molecule $H_2$ in the interstellar
medium provides a good test for possible variations of the ratio
(\ref{eq:mupe}). For example, from the study of
Ref.\cite{Reinhold06} based on comparing the $H_2$ spectral Lyman
and Werner lines observed in the Q 0347-383 and Q 0405-443 quasar
absorption systems ($z\simeq 2.6-3.0$, corresponding to look-back times
of 10-12 Gyr) with the laboratory
measurements, it was found:
\begin{equation}\label{eq:timemupe}
\frac{\dot{\mu}_{pe}}{\mupe}=(-2.16\pm 0.52)\times 10^{-15}\,{\rm
yr}^{-1}\,.
\end{equation}
This is a $\sim 4\sigma$ positive result implying a decreasing
proton mass with cosmic time evolution. According to
Eq.\,(\ref{eq:dotffp}) we have
$\frac{\dot{\mu}_{pe}}{\mupe}-(1+z)\,H(z)\frac{{\mu}'_{pe}(z)}{\mupe}$,
and the above result is thereby equivalent to an increasing proton
mass with the redshift at an approximate rate  ${\mu}'_{pe}(z)/{\mupe(z)}\sim
10^{-5}$. The result (\ref{eq:timemupe}) was questioned by other
authors\,\cite{King2008} on account of possible spectral wavelength
calibration uncertainties, rendering a significance at the
$1\,\sigma$ level only. In addition, Ref.\,\cite{Truppe2013} claims that
the bound on the ratio (\ref{eq:timemupe}) is as as tight as $\sim 10^{-17}\,{\rm yr}^{-1}$.

Future laboratory tests using atomic clocks could achieve similar
precision, if not better. If so they will verify in a robust way if that tiny
effect is there or not. The tests involve precise experiments in quantum
optics, e.g. obtained by comparing a cesium clock with 1S-2S
hydrogen transitions. In a cesium clock the time is measured using a
hyperfine transition, which is proportional to $Z\,\alpha_{\rm
em}^2(\mu_N/\mu_B)(m_e/m_p)\,R_{\infty}$, with $R_{\infty}$ the
Rydberg constant, $\mu_N$ is the nuclear magnetic moment and
$\mu_B=e\hbar/2m_pc$ is the nuclear magneton. Thus,
$\dot{\mu}_N/\mu_N\propto -\dot{m_p}/m_p\simeq -\dot{\Lambda}_{\rm
QCD}/\LQCD$. Since the hydrogen transitions are only dependent on
the electron mass (assumed constant here), we can compare the cesium
clock with hydrogen transitions over a period of time and obtain an
extremely precise atomic laboratory measurement of the ratio
(\ref{eq:mupe}). The sensitivity in the parameter variation using these techniques could lie well below the ppm limit in the future\,\cite{Rosenband08}. In the meanwhile some claims existed in the literature on  a statistically significant offset in the measurements of $\mupe$ derived on comparing terrestrial and astrophysical
microwave transitions in ammonia and other molecules, which nevertheless seem to have disappeared and turned into upper bounds of order $0.1$ ppm on the fractional variation\,\cite{Levshakov20132010}. Obviously this situation calls out for new verifications and high-precision measurements, as well as for a sound theoretical framework that can help to interpret these measurements in the context of fundamental physics.
%
%

\section{The roots of the micro and macro connection}\label{sect:seeds}

The Newtonian gravitational coupling $G$ has traditionally been
considered as a fundamental constant of Nature. Measurements of it
in laboratory experiments have more than two hundred years of
history and are constantly subject to revision. For example, the
2010 recommended value of $G$ by the CODATA group,
$G=6.67384(80)\times 10^{-11}\,{\rm m}^3\,{\rm kg}^{-1}{\rm
s}^{-2}$, is smaller than the 2006 value by the fractional amount
$6.6\times 10^{-6}$, and with a a $20\%$ increase in uncertainty of
the former compared to the latter (viz. 12 parts versus 10 in a
hundred thousand parts)\,\cite{Mohr2012}.

There is another fundamental gravitational parameter, which is about
reaching a century of rather busy (and puzzling) history, to
wit: the cosmological term $\CC$. For about fifty years it has plagued
theoretical physics with the cosmological constant problem\,\cite{Zeldovich67}, perhaps
the biggest mystery of theoretical physics ever; and whose
resolution could ultimately revolutionize the foundations of quantum
field theory and gravitation\,\cite{CCP,JSP-CCReview2013}. Both $G$
and $\CC$ are crucially involved in Einstein's field equations (in
their 1917 form\,\cite{Einstein1917}):
\begin{equation}\label{eq:EQS1}
R_{\mu \nu }-\frac{1}{2}g_{\mu \nu }R-\CC\,g_{\mu\nu}=\frac{8\pi
G}{c^4}\ {T}_{\mu\nu}\,.
\end{equation}
Here ${T}_{\mu\nu}$ is the ordinary energy-momentum tensor for
matter. Defining $\rL\,c^2=\frac{\Lambda\,c^4}{8\pi G}$ (the
so-called ``vacuum energy density'') we can move $\CC$ to the r.h.s.
of (\ref{eq:EQS1}) and rewrite Einstein's equations with a pure
geometric tensor on its l.h.s. (Einstein's tensor $G_{\mu\nu}$):
\begin{equation}\label{eq:EQS2}
G_{\mu\nu}\equiv R_{\mu \nu }-\frac{1}{2}g_{\mu \nu }R=\frac{8\pi
G}{c^4}\, \tilde{T}_{\mu\nu}\,.
\end{equation}
In the cosmological context, the quantity
\begin{equation}\label{eq:tildeTmunu}
\tilde{T}_{\mu\nu}\equiv
T_{\mu\nu}+g_{\mu\nu}\,\rL\,c^2=(\rho_{\Lambda}\,c^2-p_{m})\,g_{\mu\nu}+(\rho_{m}+p_{m}/c^2)U_{\mu}U_{\nu}
\end{equation}
is the modified energy-momentum tensor for a perfect cosmological
fluid (with $4$-velocity $U^{\mu}$) involving both the effects of
matter ($\rmr$) and vacuum energy $(\rL)$ energy densities. Notice
that the above equation implies $p_{\CC}=-\rL$ for the vacuum
``fluid'', hence negative  pressure $p_{\CC}<0$ (an
anti-gravitational effect)  for $\rL>0$.

\subsection{Vacuum dynamics and matter non-conservation}\label{sect:timemasses}

From the neatly separated form of Einstein's equations
(\ref{eq:EQS2}), in which the geometric and matter-energy  tensors
lie on different sides of the equality, one can transfer the pure
geometric identities of the Einstein tensor $G_{\mu\nu}$ into
physically measurable properties of the matter-energy ingredients
and their gravitational interactions. For example, the Bianchi
identity $\nabla^{\mu}G_{\mu\nu}=0$ implies
$\nabla^{\mu}\left(G\tilde{T}_{\mu\nu}\right)=0$. We can evaluate
the latter explicitly from Eq.\,(\ref{eq:tildeTmunu}) using the
Friedmann-Lema\^\i tre-Robertson-Walker (FLRW) metric. In doing this
we need not assume that either $G$ or $\rL$ remain constant since
the Bianchi identity does not necessarily imply it; in fact, we may
fulfil it through mutual compensations occurring among the different
parts. Assuming only time evolution of the parameters (which is
consistent with the Cosmological Principle embodied in the FLRW
metric) we find\footnote{In what follows we use natural units
$\hbar=c=1$, but shall keep explicitly Newton's constant $G=\hbar c/M_P^2\equiv 1/M_P^2$, where $M_P\simeq 1.22\times 10^{19}$ GeV is
the Planck mass.}:
\begin{equation}\label{BianchiGeneral1}
\dot{G}(\rmr+\rL)+G(\dot{\rho}_m+\dot{\rho}_{\CC})+3\,G\,H\,(\rmr+\pmr)=0\,,
\end{equation}
where the Hubble function $H=\dot{a}/a$ ($a$ being the scale factor)is related with the energy densities
through Friedmann's equation
\begin{equation}\label{eq:Friedmann}
H^2=\frac{8\pi\,G}{3}\left(\rmr+\rL\right)\,,
\end{equation}
if we assume a spatially flat Universe. The expression
(\ref{BianchiGeneral1}) can be rewritten in terms of the
cosmological redshift variable (much more convenient in cosmological
observations than the cosmic time) with the help of
Eq.(\ref{eq:dotffp}). Neglecting the matter pressure (which holds
good for the entire cosmic history after the early radiation epoch),
we obtain:
\begin{equation}\label{BianchiGeneral2}
\frac{G'(z)}{G(z)}\,\left[\rmr(z)+\rL(z)\right]+{\rho}'_m(z)+\rho'_{\CC}(z)=\frac{3\rmr(z)}{1+z}\,.
\end{equation}
In the particular case in which $G$ and $\rL$ stay strictly constant
(as in the concordance $\CC$CDM cosmology) the previous differential
equation just boils down to ${\rho}'_m(z)=3\rmr(z)/(1+z)$, and upon
integration we find $\rmr(z)=\rmo\,(1+z)^{3}$, the standard result expressing local matter
conservation. However, it is obvious that
the general local conservation equation\,(\ref{BianchiGeneral2}) offers a much wider spectrum of possibilities, and from this point of view the standard
option may be considered as a bit too restrictive. Even though we
know that the standard matter conservation law must be essentially
correct, small corrections could perhaps be accommodated. We
parameterize this possibility as follows:
\begin{equation}\label{nontandardconserv}
\rmr(z)=\rmo\,(1+z)^{3(1-\num)}\,,
\end{equation}
where $|\num|\ll 1$ is a small dimensionless parameter. At the moment, phenomenologically we can interpret Eq.\,(\ref{nontandardconserv})
either as a law for the non-conservation of the number of particles within one or more species or as a law for the non-conservation of the particle masses, or both. As in\,\cite{FritzschSola2012} we will assume that it is the particle mass of the various species that is non-conserved.

From the foregoing we find a first hint of the ``micro and macro
connection'', namely a subtle relationship  between the laws of the
subatomic world and those governing the Universe in the large that
could explain the mild, almost imperceptible, evolution of the
fundamental constants of nature. In the above context, the
tiny seeds of matter non-conservation can be consistent with the
general covariance of Einstein's equations provided the vacuum energy density and/or Newton's
gravitational coupling also evolve with the cosmological expansion,
i.e. $\rL=\rL(z)$ and/or $G=G(z)$.  For example, if we assume that
$G$ stays strictly constant and substitute (\ref{nontandardconserv})
in (\ref{BianchiGeneral2}), we find:
$\rho_{\CC}'(z)=3\,\num\,\rmo\,(1+z)^{3(1-\num)}$. Integrating and
fixing the integration constant through the condition that
$\rL(z=0)=\rLo$ (the measured value of the CC), we arrive
at the expression for the ``running'' (cf. Sect. \ref{sect:runningVacuum}) of the vacuum energy density:
\begin{equation}\label{eq:rLz}
\rL(z)=\rLo+\frac{\nuL\,\rmr^0}{1-\num}\,\left[(1+z)^{3(1-\num)}-1\right]\,,
\end{equation}
where $\nuL=\num$ in this case.  The coefficient $\nuL$ controls the running of the vacuum energy density (for $\nuL=0$ we have $\rL=\rLo$ at all times, as in the $\CC$CDM). The fact that $\nuL$ here is the same as the coefficient $\num$ that parameterizes the anomalous matter conservation law (\ref{nontandardconserv}) is especial. In the next section we will consider a more general situation where $\nuL$ and $\num$ need not be equal.

One can view it the other way around: if the vacuum energy density evolves with the expansion as in Eq.\,(\ref{eq:rLz}), the general
covariance of Einstein's equations enforces the anomalous matter
conservation law (\ref{nontandardconserv}) with $\num=\nuL$. Thus, having dynamical vacuum energy may lead to matter \textit{non}-conservation. Obviously, since
$\num$  must be small in order that the departure of
(\ref{nontandardconserv}) from the ordinary conservation law is not
too big, the ``running'' of the cosmological vacuum energy density
(\ref{eq:rLz}) should also be small ($|\nuL|\ll1$).

This is an interesting result, as it suggests that the small variation of the fundamental constants in Nature (e.g. the particle masses) could bare a dynamical relation with
the small change of the vacuum energy or cosmological constant in
the course of the cosmological expansion. In this manner the micro and
macro connection can be perfectly compatible with the (ostensibly)
constant value of the cosmological term, thus preserving the excellent
phenomenological status of the $\CC$CDM model.
But the new paradigm also says that such status is
only an approximation, a very good one indeed, but just an
approximation to the underlying dynamical character of these quantities.

\subsection{Matter non-conservation with variable $G$ and $\rL$}\label{sect:nonconservation}

Once the possibility that $\rL$ can evolve with the cosmic
time/redshift, the general local conservation equation
(\ref{BianchiGeneral2}) naturally suggest that the gravitational
coupling $G$ can be, too, time/redshift dependent.  We can easily
formulate a simple working scenario where the matter
non-conservation law (\ref{nontandardconserv}) coexists with both a
dynamical gravitational coupling $G$ and a dynamical vacuum energy
density $\rL$. Let us assume that $G$ varies logarithmically with
the Hubble function as follows:
\begin{equation}\label{GvariableH}
G(H)=\frac{G_0}{1+\nuG\,\ln\frac{H^2}{H_0^2}}\simeq G_0\,\left(1-{\nuG}\,\ln\frac{H^2}{H_0^2}+{\cal O}({\nu}_G^2)\right)\,,
\end{equation}
where $G_0=1/M_P^2$ is the current value
(i.e. for $H=H_0$) -- see \,\cite{Fossil07} for a motivation of this expression. Here ${\nuG}$ is another small dimensionless
parameter, different from $\num$ in general. The logarithmic
law (\ref{GvariableH}) is reasonable if we take into account that we
do not expect a significant variation of $G$ for very long periods
of the cosmic evolution. For ${\nuG}>0$ we have smaller $G$ in
the past (where the Universe was more energetic) and hence it
behaves as an asymptotically free coupling. Let us
substitute both expressions (\ref{nontandardconserv}) and
(\ref{GvariableH}) into the general local conservation law
(\ref{BianchiGeneral2}), then expand the resulting equations in the
small parameters $\num$ and $\nuG$ up to linear order and
neglect all terms of order ${\cal O}(\num^2)$,  ${\cal
O}(\nuG^2)$ and ${\cal O}(\num\nuG)$.  
Furthermore, using Friedmann's equation (\ref{eq:Friedmann}) and integrating the
resulting differential equation for $\rL(z)$, it is straightforward
to reach the following final result:
\begin{equation}\label{eq:rLvariableGandM}
\rL(z)=\rLo+\frac{\num+\nuG}{1-\num-\nuG}\,\rmr^0\,\left[(1+z)^{3(1-\num)}-1\right]\,,
\end{equation}
where now $\num+\nuG\equiv\nuL$ is the effective coefficient for the running of $\rL$. Eq.\,(\ref{eq:rLvariableGandM}) shows, in very concrete terms, that the dynamical character of the vacuum energy can be in interplay with both the dynamics of the gravitational coupling (if $\nuG\neq 0$) and the non-conservation of matter (if $\num\neq 0$). From (\ref{nontandardconserv}) and (\ref{eq:rLvariableGandM}) we can derive the Hubble function from Friedmann's equation (\ref{eq:Friedmann}), and we find:
\begin{equation}
{H^2(z)}=\frac{8\pi\,G}{3} \left\{\rLo+\rmo\,(1+z)^{3(1-\num)}+\frac{\nuL\rmo}{1-\nuL}\left[
(1+z)^{3(1-\num)}-1\right]\right\} \,. \label{nomalHflow}
\end{equation}

The following observations are in order:\,
\begin{itemize}
\item
i) In the particular case in which there is no cosmic running of $G$ (i.e. if $\nuG=0$),
the result (\ref{eq:rLvariableGandM}) reduces to
(\ref{eq:rLz}), as expected. Recall, however, that the latter is an exact result, whereas the former is valid at linear approximation in $\num$ and $\nuG$ (despite we have kept some higher order terms to better display the similarities between the two expressions);\,

\item
ii) If $\num=0$ Eq\,(\ref{eq:rLvariableGandM}) yields an evolution law for the vacuum energy
density very similar (in linear order) to (\ref{eq:rLz}) by just replacing
$\num$ with $\nuG$ in the latter. The evolution of $\rL$ is realized at the expense of a time-evolving Newton's coupling since now matter is strictly conserved;\,

\item
iii) If $\nuG=-\num$, we have $\nuL=0$ and the vacuum energy density remains static at all times, $\rL=\rLo$,  as in the $\CC$CDM. However, the anomalous matter conservation law
(\ref{nontandardconserv}) still holds thanks to the running $G$, Eq.\, (\ref{GvariableH})\footnote{The function $G(H)$ satisfying
Eq.\,(\ref{BianchiGeneral2}) at fixed $\rL$ and with a
matter energy density evolving as in (\ref{nontandardconserv}) is given by
$G/G_0=\left(H^2/H_0^2\right)^{\num}$. This result is exact, it does not depend on $\num$ being small. However, expanding it for small
$\num$ we recover Eq.\,(\ref{GvariableH}) for
$\num=-\nuG$, as we should.}.

\end{itemize}

Notice that the scenario ii) above suggests that, if there is strict
matter conservation ($\num=0$), it is still possible to have a variable vacuum energy density as in Eq.\,(\ref{eq:rLvariableGandM}),
provided the gravitational coupling is logarithmically running (i.e. with $\nuG$ small, but nonvanishing).
On the other hand, scenario iii) says that in order to have the anomalous
matter conservation (\ref{nontandardconserv}) it is not indispensable to have dynamical vacuum
energy, inasmuch as the Newtonian coupling evolves logarithmically with the expansion rate. Since a mild logarithmic evolution cannot be excluded, these possibilities should be taken seriously into account.

Beyond particular implementations, however, the main message from the previous considerations is that in general we should \textit{not} expect strict matter conservation in an accelerating Universe. In the next section we address an specific theoretical context.

\section{Running vacuum energy in quantum field theory}\label{sect:runningVacuum}

The following question is  pertinent and even crucial: is it
conceivable the possibility of having running vacuum energy, say as in (\ref{eq:rLvariableGandM}), in fundamental physics?  The vacuum
energy-density of an expanding universe is expected to change with time,
and one may conceive theoretical proposals supporting this
possibility. These originate from quantum field theory (QFT) in
curved spacetime, see \,\cite{JSP-CCReview2013} and references therein\,\footnote{See also \cite{Terazawa2012,Bennie} for alternative formulations, and \cite{Tommi2014,Maroto2014,LBS,CliftonBarrow2014} for recent developments.}.  Not only so, some of these QFT-inspired models have been recently put to
the test\,\cite{GoSol2014,GoSolBas2014,Bas2015} and one
finds they are able to pass all the observational tests in a way
comparable to the standard $\CC$CDM model. This is
an interesting feature because a dynamical vacuum model in which
$\rL=\rL(t)$ could better help explaining the phase transitions of
the Universe at different stages and the large entropy problem \,\cite{SolGoReview2015}.

It is well-known that in particle physics we have theories such as
QED or QCD  where the corresponding gauge coupling constants $g_i$
run with an energy scale $\mu_R$, i.e. $g_i=g_i(\mu_R)$. Along the same lines, the following form has been proposed for
the renormalization group (RG) equation for the vacuum energy
density of the expanding
Universe (cf. \,\cite{JSP-CCReview2013}, \cite{SolGoReview2015} and references therein):
\begin{equation}\label{seriesLambda}
\frac{d\rL(\mu_c)}{d\ln\mu_c^2}=\frac{1}{(4\pi)^2}\left[\sum_{i}\,B_{i}M_{i}^{2}\,\mu_c^{2}
+\sum_{i}
\,C_{i}\,\mu_c^{4}+\sum_{i}\frac{\,D_{i}}{M_{i}^{2}}\,\mu_c^{6}\,\,+...\right]\,.
\end{equation}
In this expression, $M_{i}$ are the masses of the particles
contributing in the loops, and $B_{i},C_i,..$ are dimensionless
parameters. The RG equation (\ref{seriesLambda}) provides the rate
of change of the quantum effects on $\rL$ as a function of the cosmic
scale $\mu_c$.

The energy scale $\mu_c$ should naturally be associated to a cosmic energy variable. In the FLRW metric we naturally expect  $\mu_c=H$.
Integration provides the following leading expression:
\begin{equation}\label{lambdaH2H4}
\rL(H) = \frac{3}{8\pi G}\,\left(c_0 + \nuL H^{2} +
\frac{H^{4}}{H_{I}^{2}}\right) \;,
\end{equation}
where $c_0$ has dimension $2$ (in natural units) and we have introduced the
dimensionless coefficient $\nuL$ and the dimensionful one $H_I$.
Comparing with (\ref{seriesLambda}) it is easy to see that
\begin{equation}\label{eq:nualphaloopcoeff}
\nuL=\frac{1}{6\pi}\, \sum_{i=f,b} B_i\frac{M_i^2}{M_P^2}\,.
\end{equation}
This coefficient is related to the $\beta$-function that controls the running of $\rL$. The sum
in (\ref{eq:nualphaloopcoeff}) involves both fermions and bosons ($i=f,b$)-- see e.g.
\cite{Fossil07} for a one-loop calculation.
The $\sim H^4$ term in (\ref{lambdaH2H4}) can be used to describe inflation in the early universe\,\cite{SolGoReview2015,LBS} and the dimensionful coefficient $H_I$ represents the scale of inflation. In what follows we shall consider only the effects in the post-inflationary universe and hence we can focus only on the first two terms of (\ref{lambdaH2H4}) since $H^4$ is negligible now. For our purposes, therefore, it suffices to consider the expression
\begin{equation}\label{lambdaTypeA1}
\rL(H) = \frac{3}{8\pi G}\,\left(c_0 + \nuL H^{2}\right)\,.
\end{equation}
The remarkable feature to stand out at this point is the following: the class of running vacuum energy densities of the form (\ref{lambdaTypeA1}) -- in which $G$ can be variable -- leads to the following exact expression for the Hubble function:
\begin{equation}
H^2(t)=\frac{8\pi\,G(t)}{3} \left\{\rLo+\rmr(t)+\frac{\nuL}{1-\nuL}\left[\rmr(t)-\rmo
\right]\right\} \,, \label{eq:exact}
\end{equation}
where use has been made of Friedmann's Eq.\,(\ref{eq:Friedmann}). In particular, if $G$ is running as in (\ref{GvariableH}) and the anomalous matter conservation law for $\rmr$ is as in (\ref{nontandardconserv}),  we are immediately led to Eq.\,(\ref{nomalHflow}) and (a fortiori) to Eq.\,(\ref{eq:rLvariableGandM}) --- with the identification $\nuL=\num+\nuG$. It means that our entire dynamical framework (and with it the roots of the micro and macro connection) can ultimately be derived from the RG-equation (\ref{seriesLambda}). There is an even larger class of vacuum models providing similar scenarios -- see Ref.\,\cite{GoSol2014,GoSolBas2014} for detailed analyses, and \cite{SolGoReview2015} for additional implications.

\section{The cosmic running of the particle masses and couplings }

The  matter density of the universe can be approximated as follows:
$
\rM\simeq n_p\,m_p+n_n\,m_n+n_X\,m_X\,,
$
where we neglect the leptonic contribution and the relativistic
component (photons and neutrinos).
Here $n_p, n_n, n_X\, (m_p,m_n,m_X)$ are the number densities
(and masses) of protons, neutrons and dark matter (DM)
particles, respectively. Assuming that the mass non-conservation law alluded to in Sect\,\ref{sect:seeds} is to be attributed to the change of the mass of the particles, the total rate of change of the mass density associated to such mass anomaly can be estimated
as follows\, \cite{FritzschSola2012}:
\begin{equation}\label{eq:reltimeMdensityUniv}
\frac{\delta\dot{\rho}_m}{\rM}\simeq\left(1-\frac{\OMB}{\ODM}\right)\left(\frac{\OMB}{\ODM}\,\frac{\dot{m}_B}{m_B}+\frac{\dot{m}_X}{m_X}\right)\,,
\end{equation}
where we have set $m_n=m_p\equiv m_B$ and used the cosmological parameters  $\OMBo$ and $\ODMo$ for baryons and  DM, respectively. We note that the leading contribution from $n_n/n_p$ appears at second order and can be neglected since  $n_n/n_p\sim 10\%$ after the primordial nucleosynthesis. At the same time, with the help of Eq.\,(\ref{nontandardconserv}) we find  that the \textit{l.h.s.} of (\ref{eq:reltimeMdensityUniv}) reads  ${\delta\dot{\rho}_m}/{\rM}\simeq 3\num\,H$, in good approximation. The corresponding cosmic drift rates of the vacuum energy density and gravitational coupling ensue from  (\ref{eq:rLvariableGandM}) and  (\ref{GvariableH}), respectively:
\begin{equation}\label{eq:deltaLambdaG}
\frac{\dot{\rho}_{\CC}}{\rL}\simeq -3\left(\num+\nuG\right)\,\frac{\OMo}{\OLo}\,(1+z)^3\,H\,, \ \ \ \ \ \frac{\dot{G}}{G}=-2\nu_G\frac{\dot{H}}{H}\,,
\end{equation}
where $\dot{H}$ can be computed from (\ref{nomalHflow}) using once more Eq.\,(\ref{eq:dotffp}).

We define $\nueff={\num}/(1-\OMBo/\ODMo)$ and introduce  the anomaly indices $\nuB$ and $\nuX$ for baryon and DM mass non-conservation\,\footnote{We are generalizing here the results of \cite{FritzschSola2012} for the case when both $\rL$ and $G$ are running at the expense of matter non-conservation. Note that the definition of $\nu_B$  has changed now as compared to that reference.}:
\begin{equation}\label{eq:rateNandDM}
\frac{\dot{m}_B}{m_B}\simeq\frac{\dot{\Lambda}_{\rm QCD}}{\LQCD}=3\,\nu_B\,H\,,\ \ \ \ \ \ \ \ \frac{\dot{m}_{X}}{m_X}=3\,\nuX\,H\,,
\end{equation}
where we may approximately set $m_B\propto\,\LQCD$ from (\ref{eq:ProtonMass}). Mind that because of (\ref{eq:reltimeMdensityUniv}) the mass anomaly indices are constrained by the relation $\nueff=\left({\OMBo}/{\ODMo}\right)\nu_B+\nuX$.
Using (\ref{eq:dotffp}) and integrating, we find the redshift evolution of the masses:
\begin{equation}\label{eq:LQCDmxDz}
m_i(z)=m_i^0\,\left(1+z\right)^{-3\,\nu_i}\, \ \ \Rightarrow\ \ \ \frac{\delta m_i(z)}{m_i}\simeq -3\nu_i\,\ln(1+z)\,,
\end{equation}
for baryon and DM particles $i=B,X$,  and an entirely similar formula for $\LQCD$ to that for $m_B$. We have defined $\delta m_i(z)=m_i(z)-m_i^0$, where $m_i^0\equiv m_i(z=0)$ are the current masses of these particles.  Their cosmic evolution is very mild since it is logarithmic with the redshift and $|\nu_i|\ll 1$.

The previous equations lead also to the redshift evolution of
the masses of all chemical elements in the universe. For a nucleus of atomic number $A$ and current mass $M_A^0\simeq A\,m_B^0-B_A$ we can neglect the cosmic shift on the binding energy,  $B_A$, since at leading order
relies on pion exchange among the nucleons, and the pion
mass has a softer dependence on $\LQCD$: $m_{\pi}\sim \sqrt{m_q\,\LQCD}$,
due to the chiral symmetry. Thus, at leading order, we obtain:
\begin{equation}\label{eq:MAz}
M_A(z)\simeq A\,m_B^0\,\left(1+z\right)^{-3\,\nu_B}-B_A\,.
\end{equation}
Notice that although the chemical elements change their masses, a disappearance or
overproduction of nuclear mass in the universe (depending on the sign of $\nu_B$) is
compensated for by the running of the vacumm energy $\rL$ and/or the running of $G$, see Eq.\,(\ref{eq:deltaLambdaG}).
In particular, if $\nuG=0$ the signs of $\delta\dot{\rho}_m$ and $\dot{\rho}_{\CC}$ are opposite, as it should be expected. Similarly, if $\num=0$ (entailing matter conservation), we find from (\ref{eq:deltaLambdaG}) that the signs of $\dot{\rho}_{\CC}$ and $\dot{G}$ are also opposite (since $\dot{H}<0$), as also expected. In both cases the feedback is a direct reflex of the Bianchi identity and hence is fully consistent with the general covariance of the theory.

Let us express the cosmic drift of masses in terms of the Hubble function, which acts as the original running scale in cosmology, $\mu_c=H$ (cf. Sect.\,\ref{sect:runningVacuum}). Using (\ref{nomalHflow}) and after a straightforward calculation we find, at leading order ($i=B,X$):
\begin{equation}\label{eq:LQCDH}
\frac{\delta m_i(H)}{m_i}\simeq-\frac{\nu_i}{1-\num}\,\ln\left[\frac{G_0}{G}\frac{1-\nuL}{\OMo}\,\frac{H^2}{H_0^2}-\frac{\OLo-\nuL}{\OMo}\right]\,,
\end{equation}
where $G=G(H)$.
In the previous formula  $\OMo=\OMBo+\ODMo$ and we have defined  $\delta m_i(H)=m_i(H)-m_i^0$.
We recall from Sect.\,\ref{sect:nonconservation} that  $\nuL=\num+\nuG$. The corresponding  mass variation of a nucleus, $\delta M_A(H)$, can also be obtained easily from (\ref{eq:MAz}) upon using the above prescriptions.

Finally, we compute the cosmic evolution of the QCD gauge coupling (\ref{alphasLQCD}) in terms of the Hubble function. The result reads:
\begin{equation}\label{eq:alphasH}
\frac{\delta\alpha_s(\mu_R;H)}{\alpha_s}\simeq-\frac{b}{2\pi}\,
\frac{\nu_B}{1-\num}\,\ln{\left[\frac{G_0}{G}\frac{1-\nuL}{\OMo}\,\frac{H^2}{H_0^2}-\frac{\OLo-\nuL}{\OMo}\right]}\,.
\end{equation}
Here $\delta\alpha_s(\mu_R;H)=\alpha_s(\mu_R;H)-\alpha_s(\mu_R;H_0)$ is the difference, at fixed $\mu_R$, between the values of $\alpha_s$  at the cosmic epoch when the Hubble function was  $H(t)$ and at present ($H(t_0)=H_0$).  We observe that the strong coupling became a function of \textit{two} running scales, $\mu_R$ and $\mu_c=H$. Since $b>0$ in QCD, it turns out that for $\nu_B>0$ the strong coupling $\alpha_s(\mu_R;H)$ is ``doubly asymptotically free'', namely it decreases for large $\mu_R$ and also for large $H$ (meaning that in the past it was smaller than it is at present), whereas for $\nu_B<0$ the cosmic evolution drives the running of $\alpha_s$
opposite to the normal QCD running.

Of course the running of the strong coupling with $\mu_c=H$ is much slower than the ordinary gauge running with $\mu_R$.
Detecting the former, however, is the main target of the experiments that aim at finding evidence of the time variation of fundamental constants with the aid of atomic clocks and astrophysical observations. Here such cosmic time variation of the ``fundamental constants'' has just been rephrased on quantum field theoretical terms\footnote{The time evolution of the various particle masses and gravitational parameters described here  can be equivalently formulated, if desired, entirely in terms of dimensionless quantities such as e.g. $\CC/m_p^2$ and $Gm_p^2$, see \cite{FritzschSola2012}, and it is therefore not an artifact of starting from dimensional quantities. This leads to new sources of violation of Einstein's Equivalence Principle, different from other known mechanisms\,\cite{Damour2012}. }.


\section{Conclusions}

In this work we have considered a possible origin for the cosmic time variation of the fundamental constants, such as particle masses and couplings. Usually one tries to confine the possible explanation of such variation within the strict domain of particle physics and in particular of grand unified theories of electroweak and strong interactions. However, if one includes the gravitational coupling and the cosmological constant (or vacuum energy density $\rL$), the explanation can be given in a wider context. We have called this context the ``micro and macro connection''.
By admitting a mild dynamical behavior of the vacuum energy density, $\rL=\rL(t)$, and/or of Newtons's coupling, $G=G(t)$, we find that the time drift of the $\LQCD$ parameter and of the masses of all the baryons and dark matter particles in the universe, can be correlated with the cosmic change of the gravitational parameters. Formally this is attained by preserving the Bianchi identity and hence ultimately hinges on the fundamental principle of covariance of General Relativity. The small change of the gravitational pair $(\rL,G)$ through the cosmic evolution can be compensated for by a corresponding change of the couplings and of the particle masses, not only of baryons but also of the the dark matter particles. The current limits on the variation of the particle masses and those of the cosmological parameters are compatible and make such micro and macro connection, in principle, viable and theoretically appealing.

\vspace{0.5cm}

{\bf Acknowledgments}\hspace{0.3cm} The work of JS has been partially supported by
FPA2013-46570 (MICINN), Consolider grant CSD2007-00042 (CPAN) and by
2014-SGR-104 (Generalitat de Catalunya). We are grateful to the Institute for Advanced Study at the Nanyang Technological University in Singapore for hospitality and support.

\newcommand{\JHEP}[3]{ {JHEP} {#1} (#2)  {#3}}
\newcommand{\NPB}[3]{{ Nucl. Phys. } {\bf B#1} (#2)  {#3}}
\newcommand{\NPPS}[3]{{ Nucl. Phys. Proc. Supp. } {\bf #1} (#2)  {#3}}
\newcommand{\PRD}[3]{{ Phys. Rev. } {\bf D#1} (#2)   {#3}}
\newcommand{\PLB}[3]{{ Phys. Lett. } {\bf B#1} (#2)  {#3}}
\newcommand{\EPJ}[3]{{ Eur. Phys. J } {\bf C#1} (#2)  {#3}}
\newcommand{\PR}[3]{{ Phys. Rep. } {\bf #1} (#2)  {#3}}
\newcommand{\RMP}[3]{{ Rev. Mod. Phys. } {\bf #1} (#2)  {#3}}
\newcommand{\IJMP}[3]{{ Int. J. of Mod. Phys. } {\bf #1} (#2)  {#3}}
\newcommand{\PRL}[3]{{ Phys. Rev. Lett. } {\bf #1} (#2) {#3}}
\newcommand{\ZFP}[3]{{ Zeitsch. f. Physik } {\bf C#1} (#2)  {#3}}
\newcommand{\MPLA}[3]{{ Mod. Phys. Lett. } {\bf A#1} (#2) {#3}}
\newcommand{\CQG}[3]{{ Class. Quant. Grav. } {\bf #1} (#2) {#3}}
\newcommand{\JCAP}[3]{{ JCAP} {\bf#1} (#2)  {#3}}
\newcommand{\APJ}[3]{{ Astrophys. J. } {\bf #1} (#2)  {#3}}
\newcommand{\AMJ}[3]{{ Astronom. J. } {\bf #1} (#2)  {#3}}
\newcommand{\APP}[3]{{ Astropart. Phys. } {\bf #1} (#2)  {#3}}
\newcommand{\AAP}[3]{{ A\&A } {\bf #1} (#2)  {#3}}
\newcommand{\MNRAS}[3]{{ Mon. Not. Roy. Astron. Soc.} {\bf #1} (#2)  {#3}}
\newcommand{\JPA}[3]{{ J. Phys. A: Math. Theor.} {\bf #1} (#2)  {#3}}
\newcommand{\ProgS}[3]{{ Prog. Theor. Phys. Supp.} {\bf #1} (#2)  {#3}}
\newcommand{\APJS}[3]{{ Astrophys. J. Supl.} {\bf #1} (#2)  {#3}}

\newcommand{\Prog}[3]{{ Prog. Theor. Phys.} {\bf #1}  (#2) {#3}}
\newcommand{\IJMPA}[3]{{ Int. J. of Mod. Phys. A} {\bf #1}  {(#2)} {#3}}
\newcommand{\IJMPD}[3]{{ Int. J. of Mod. Phys. D} {\bf #1}  {(#2)} {#3}}
\newcommand{\GRG}[3]{{ Gen. Rel. Grav.} {\bf #1}  {(#2)} {#3}}


\end{document}